\documentstyle[aas2pp4,12pt]{article}

\def\mm{M\'endez }

\begin{document}

\title{Lag of Low-Energy Photons in an X-ray Burst Oscillation: Doppler Delays}

\author{Eric C. Ford}

\affil{Astronomical Institute, ``Anton Pannekoek'',
University of Amsterdam, Kruislaan 403, 1098 SJ Amsterdam,
The Netherlands}

\authoremail{ecford@astro.uva.nl}

\bigskip
\centerline{\bf {accepted ApJ Letters, 27 April 1999}}

\begin{abstract}

Numerous X-ray bursts show strong oscillations in their flux at
several hundred Hz as revealed by RXTE. Analyzing one such oscillation
from the X-ray binary Aql X-1, I find that low energy photons
(3.5--5.7 keV) lag high energy photons ($>5.7$ keV) by approximately 1
radian. The oscillations are thought to be produced by hot spots on
the spinning neutron star. The lags can then be explained by a Doppler
shifting of emission from the hot spots; higher energy photons being
emitted earlier in the spin phase as the spot approaches the
observer. A quantitative test of this simple model shows a remarkable
agreement with the data. Similar low energy lags have been measured in
kilohertz quasi-periodic oscillations and in the accreting millisecond
pulsar SAX J1808.4-3658. A Doppler delay mechanism may be at work
there as well.

\end{abstract}

\keywords{accretion, accretion disks --- black holes -- stars:
neutron --- X--rays: stars}

\section{Introduction}

The Rossi X-Ray Timing Explorer (RXTE) has uncovered strong
oscillations of X-ray flux during X-ray bursts in several low mass
X-ray binaries (Strohmayer et al. 1996).  Current interpretation
favors a rotation mechanism for the burst oscillations: asymmetric
nuclear burning leaves a `hot spot' which rotates with the neutron
star and produces a strong modulation (Strohmayer et al. 1998).  The
frequency of the burst oscillation is then the spin frequency of the
neutron star, or twice the spin frequency for two spots (Miller 1999).
Oscillations have been discovered in X-ray bursts from the following
systems: 4U 1728-34 (363 Hz, Strohmayer et al. 1996; Strohmayer, Zhang
\& Swank 1997), KS~1731-260 (524 Hz, Smith, Morgan \& Bradt 1996), a
source near the galactic center (589 Hz, Strohmayer, Jahoda \& Giles
1997), Aql X-1 (549 Hz, Zhang et al. 1998), 4U 1636-536 (581/290 Hz,
Strohmayer et al. 1998, Miller 1999), and 4U 1702-429 (330 Hz,
Markwardt, Strohmayer \& Swank 1999). The observed frequencies are
close to the 401 Hz spin frequency of the accreting millisecond pulsar
SAX J1808.4-3658 (Wijnands \& van der Klis 1998), further
strengthening the identification of these frequencies with the neutron
star spin.

The detailed energy dependence of these burst oscillations is one
avenue that remains to be explored. Here I show that the low energy
photons in a burst oscillation from Aql X-1 lag the high energy
photons by roughly 15\% of the oscillation period. Lags of the same
sign and similar magnitudes have also been detected in other fast
signals from low mass X-ray binaries: the kilohertz quasi-periodic
oscillations, QPOs (Vaughan et al. 1997, 1998, Kaaret et al. 1999) and
the SAX J1808.4-3658 pulsed emission (Cui, Morgan \& Titarchuk 1998).

A simple mechanism of Doppler shifted emission may explain these lags.
Strong Doppler effects are expected to be important since the fast
spin rates imply high speeds ($\beta=v/c \sim0.1$).  As a hot spot on
the spinning neutron star approaches the observer (at early phases)
the emission is Doppler boosted and blue shifted, as it recedes (at
later phases) the emission is deboosted and red shifted. At early
phases the spectra are also attenuated due to the smaller projected
area.  The result is that low energy photons are preferentially
emitted after the high energy photons. A quantitative test of this
Doppler delay scenario matches the observed low energy lags in Aql~X-1
well. The possibility of Doppler effects and the fact that they may
manifest in pulse phase spectroscopy has been noted before by
Strohmayer et al. (1998).

In the next section I present the measurement of the lag in the X-ray
burst from Aql X-1. In Section~3 I describe a simple model for the
relativistic effects and compare the predicted delays to those
observed. Section~4 discusses these results in a broader context.

\section{Measurements}

For this analysis I consider the X-ray burst from Aql~X-1 starting 1
1997 March 1 23:27:40 UTC (see Zhang et al. 1998 for a report of this
burst). I use data from the RXTE Proportional Counter Array (PCA) in
an `event' mode with high time resolution (122$\mu$sec) and high
energy resolution (64 channels). A section of the lightcurve is shown
in Figure~1 (top). There are gaps in the event mode data since the
required telemetry rate is high.  Within the 4 second time window
shown in Figure~1 (top), the power density spectrum for all the
channels shows a strong oscillation at 549.7 Hz (Figure~1, bottom). In
the following I calculate Fourier transforms within this time window.

Phase delays in a signal between two energy bands are quantified by
means of cross spectral analysis (van der Klis et al. 1987; for more
information see Vaughan et al. 1994; Nowak et al. 1999).  The cross
spectrum is defined as $C(j) = X_1^{*}(j) X_2(j)$, where $X$ are the
measured complex Fourier coefficients for the two energy bands at a
frequency $\nu_j$. The phase lag between the signals in the two bands
is given by the argument of $C$ (its position angle in the complex
plane). The error in the phase lag is calculated here from the
coherence function uncorrected for counting statistics (Nowak et
al. 1999).  The cross correlation code used here has been employed to
calculate phase lags in black hole candidates (Ford et al. 1999), SAX
J1808.4-3658 and kilohertz QPOs and matches the results reported in
the literature.

Figure~2 shows the resulting phase lags from the cross spectra of the
4 seconds of data described above. Negative numbers indicate that the
oscillations in the low energy band (3.5--5.7 keV) lag those in the
higher energy bands. The lags are calculated by averaging the signal
in the range 549.6--550.1 Hz. The delays in each band up to 30 keV
(where background dominates) are 3$\sigma$ significant. The delay
between 3.5--5.7 keV and the entire 5.7--43.6 keV band is
$0.93\pm0.18$ rad, 5$\sigma$ significant.

Deadtime effects can in principle affect the measured phase lag.  The
data considered here are in the tail of the burst (rate of 9280 c/s,
full energy band) where deadtime is less important.  One method of
correcting for deadtime is to subtract a cross vector averaged over
high frequencies where no correlation is expected (van der Klis et
al. 1987). Employing this correction does not change the values
measured here.

Due to the data gaps it is not possible to perform cross correlations
on long stretches of data earlier in the burst. Cross correlations on
0.5 sec intervals of data earlier in the burst return large errors on
phase delays with inconclusive results.

\section{Model}

As a simple model for the lags I consider discrete hot spots on the
surface of the rotating neutron star. The rest frame emission of the
clump is a blackbody.  The observed spectrum at frequency $\nu$ at
spin phase $\theta$ is:
$$ F_\nu(\nu) = A_0 cos\delta [\gamma^{-1} (1-\beta\mu
cos\theta)^{-1}]^3 $$
$$~~~~~~~~~ \times ~~ \nu^3 [exp(\nu/kT)-1]^{-1} $$ where $\beta=v/c$,
$\gamma$ is the Lorentz factor, $kT=kT_0 \gamma^{-1} (1-\beta\mu
cos\theta)^{-1}$ (with $kT_0$ the rest frame temperature), $\mu$ is
the sine of the angle between the spin axis and the line of sight, and
$A_0$ is a normalization. The above formula is a relativistic
transformation of the blackbody which shifts $kT$ and modifies the
normalization such that $F/ \nu^3$ is conserved (see Rybicki \&
Lightman 1979).  The $cos\delta$ term is an area projection factor,
with $\delta$ the angle between the normal and line of sight in the
rest frame ($\delta \sim \pi-\theta$).  The phase angle, $\theta$, is
defined such that phase zero is with the spot approaching the observer
directly. The spots are considered small and isotropically emitting in
the rest frame.

I take $kT_0$=1 keV, $\beta=0.1$. These are values appropriate for the
neutron star; a more exact value of $kT_0$ is in principle possible
from the spectral fits but this depends on the fraction of the surface
contributing to the modulated hot spot emission. A more exact value of
$\beta$ depends on the neutron star radius.  I also take $\mu=1$,
i.e. a line of sight through the equator.  The spin frequency is 275
Hz and two antipodal hot spots produce an oscillation at 550 Hz. Such
a geometry, where the $\sim$550 Hz signal is a harmonic of the spin,
is suggested by recent results on other burst oscillations (Miller
1999).  The resulting spectra are blackbodies whose temperature shifts
by 10\% over the period. Averaged over phase, the spectrum is
approximately blackbody in shape with $kT$ within 1\% of the input
$kT_0$.

The spectra as a function of $\theta$, folded through the RXTE
response matrix, yield lightcurves of count rates in various energy
bands. From these lightcurves I calculate the phase lag in the 550 Hz
signal with the FFT and cross-correlation programs used in the
measurements above. The results of this calculation are shown with the
data in Figure~2.  There are no free parameters, only the assumptions
taken above.  The calculated lags will decrease if $kT_0$ is increased
or $\beta$ is decreased.  The smaller delay for higher $kT_0$ happens
since the peak of the lightcurves comes later in phase for higher
energy photons, corresponding to a smaller delay between high and low
energy photons.  Observing at higher inclinations (decreased $\mu$)
will also decrease the lag. The lightcurves that yield these predicted
lags generally have maxima at earlier phases for higher energies and
are more sharply peaked in shape at higher energies.

This simple model neglects general relativistic effects
(e.g. Strohmayer 1992, Miller \& Lamb 1998).  Two main factors from GR
will effect the observed light curves. Gravitational bending makes the
spots observable at $\theta<0$ or $\theta>\pi$, stretching the
pulse. Light travel time delays, longer for more extreme bending, will
also shift the pulse. These effects depend on the compactness of the
star. Given the quality of the present data, a more detailed treatment
including these effects is not justified.  An overall gravitational
redshift also means that $kT$ in the local frame is higher, as in
X-ray burst spectral models.

\section{Discussion}

The previous sections show that low energy photons lag high energy
photons in the oscillation signal of an X-ray burst from Aql X-1. The
sign and magnitude of the lags are in agreement with the simple model
considered in Section~3 of two hot spots on the neutron star producing
a Doppler boosted and shifted spectra as the star rotates.

This Doppler delay mechanism for producing low energy lags may
describe not only the lags in the X-ray burst oscillations but also
the lags in the accreting millisecond pulsar SAX J1808.4-3658 (Cui,
Morgan \& Titarchuk 1998) and the (lower frequency) kilohertz QPOs
(Vaughan et al. 1997, 1998; Kaaret et al. 1999). Both show a lag of
low energy photons relative to high energy photons with magnitudes of
roughly $\sim$100 $\mu$sec ($\sim0.3$ rad) for SAX J1808.4-3658 and
$\sim$30 $\mu$sec ($\sim0.2$ rad) for the kilohertz QPOs in similar
energy bands to those considered here. Some models link the frequency
of the kilohertz QPOs to a Keplerian motion in the disk (Miller, Lamb
\& Psaltis 1998, Stella \& Vietri 1999; but see Titarchuk, Lapidus \&
Muslimov 1998).  If any of the kilohertz QPOs is a result of Keplerian
motion, one might expect a soft lag due to Doppler delays.  Such lags
have been observed in what is likely the lower frequency of the two
QPOs.

Doppler delays are an alternative to previous mechanisms invoked to
produce lags.  Comptonization has been one process used to explain low
energy lags in SAX J1808.4-3658 (Cui, Morgan \& Titarchuk 1998). Low
energy lags are produced if high energy photons are injected into a
relatively cool Comptonizing cloud. This is the opposite of the
situation normally considered: Comptonization by a hot cloud in the
same region. A hot cloud produces a lag of high energy photons, as
shown quantitatively for fast signals by Lee \& Miller (1998).
Another mechanism suggested for low energy delays is an extended,
cooling hot spot with lower energy photons from the outer regions
(Cui, Morgan \& Titarchuk 1998).

More measurements of phase lags in X-ray burst oscillations are
clearly needed, in particular in the $\sim$350 Hz oscillations which
are likely from single spots. Improved statistics will also yield a
better test of the predicted energy dependence of the lags.

I thank Michiel van der Klis, Jan van Paradijs, Mariano \mm and Walter
Lewin for helpful comments.  I thank Katja Pottschmidt and coworkers
at the University of Tuebingen for comparisons of our cross
correlation codes.  I acknowledge support by the Netherlands
Foundation for Research in Astronomy with financial aid from the
Netherlands Organization for Scientific Research (NWO) under contract
numbers 782-376-011 and 781-76-017 and by the Netherlands
Researchschool for Astronomy (NOVA).


\begin{figure*}
\figurenum{1}
\epsscale{1.7}
\plotone{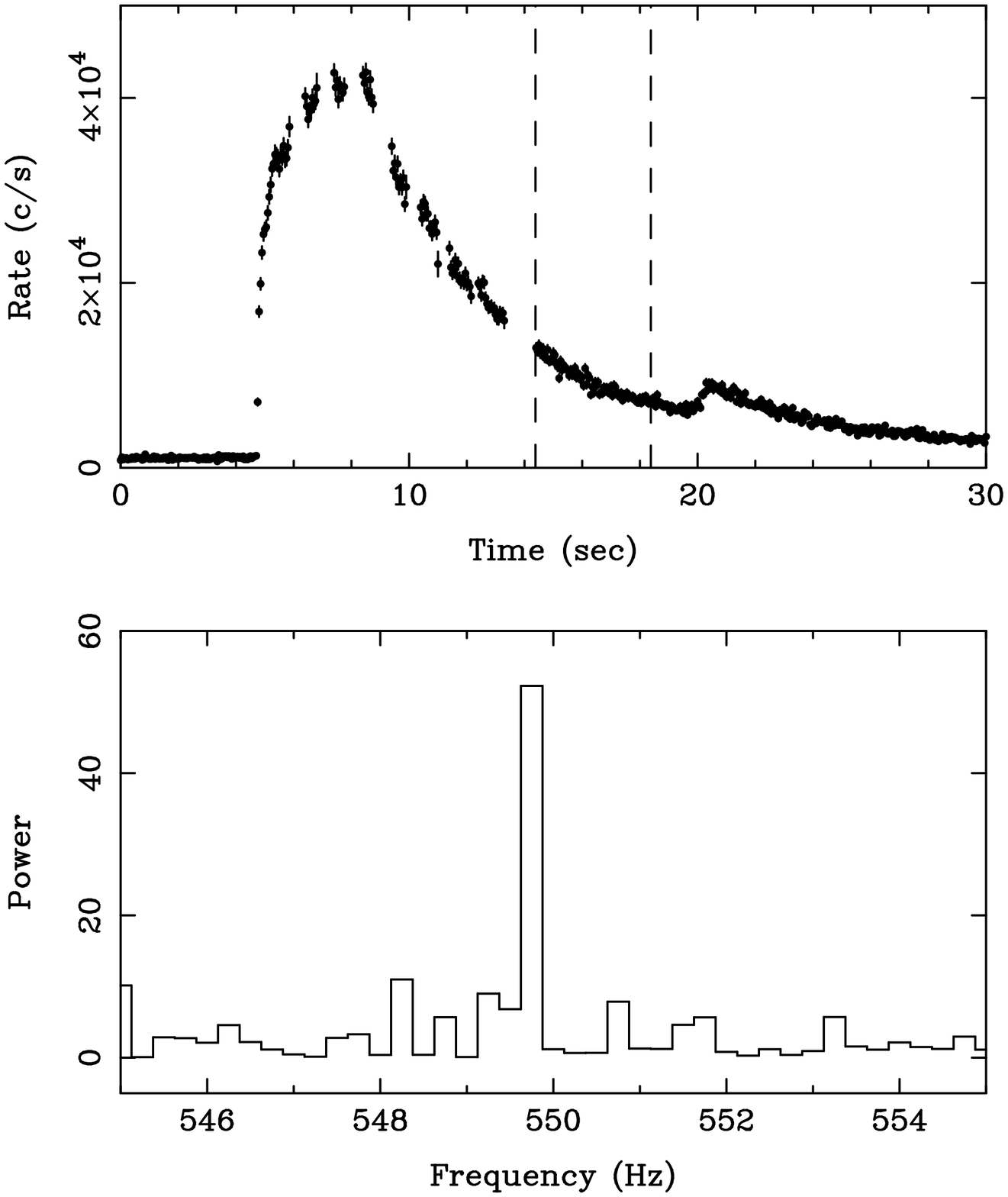} 
\caption{The lightcurve (top) and power spectrum (bottom) for the 1997
March 1 X-ray burst from Aql X-1. The power spectrum is from four
seconds of data taken in the time window shown by the dashed
lines. Data gaps are due to telemetry overloads in this PCA mode.}
\label{fig:lcpds}
\end{figure*}

\begin{figure*}
\figurenum{2}
\epsscale{1.7}
\plotone{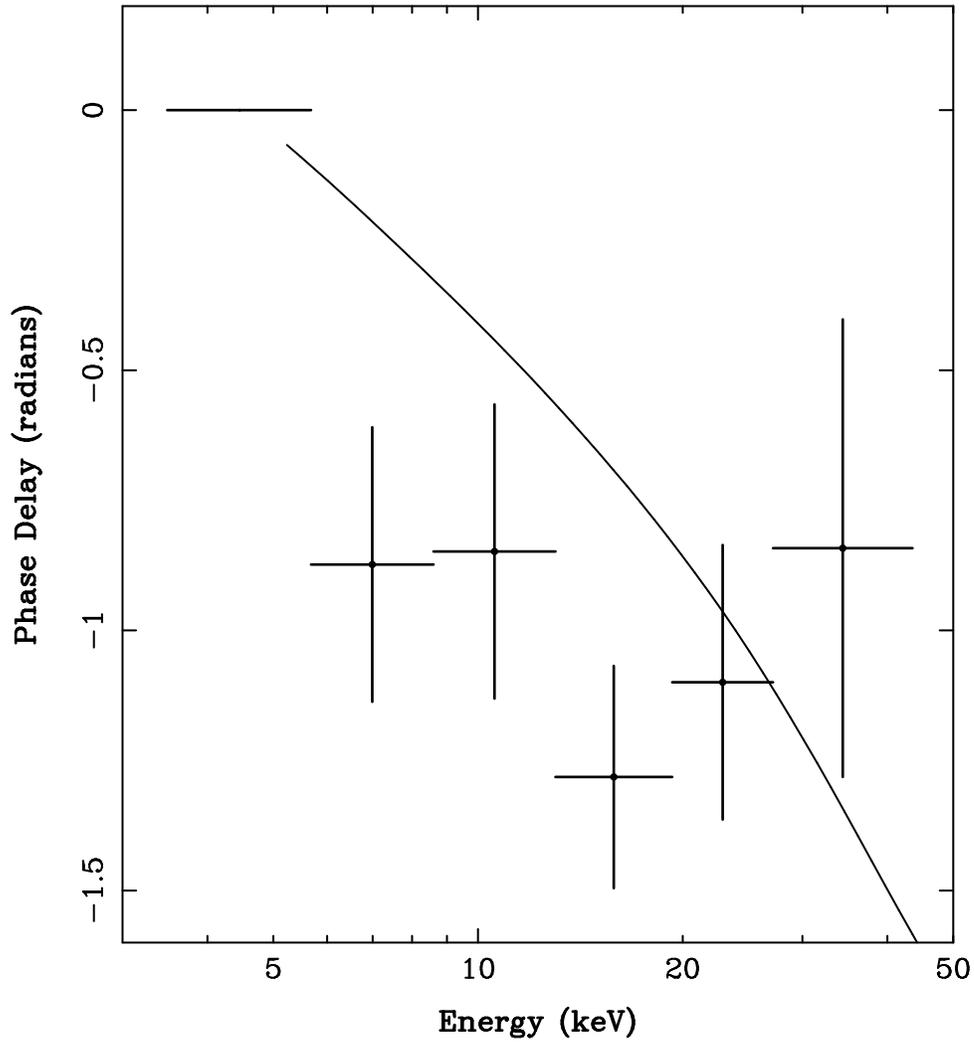} 
\caption{Phase delay measurements for the X-ray burst oscillation shown in
Figure~1 relative to the 3.7--5.7 keV band. A negative value indicates
that low energy photons lag high energy photons. The solid line is the
Doppler delay model for two hot spots (see text).}
\label{fig:lags}
\end{figure*}

\end{document}